\documentclass{article}

\usepackage[dvips]{epsfig}

\input{epsf}

\begin{document}

\title{
{\noindent\small UNITU-THEP-11/03 \hfill HD-THEP-03-42}\\ ~\\
The Kugo--Ojima confinement criterion and\\ 
the Infrared Behavior of Landau gauge QCD
\footnote{Based on an invited talk given by R.A.\  at the Second International 
Conference ``Nuclear and Particle physics with CEBAF at Jefferson
Laboratory, Dubrovnik, Croatia, 26 -- 31 May, 2003}
}

\author{Reinhard Alkofer$^a$ and Christian S.\ Fischer$^{a,b}$\\[5pt]
         \small \it $^a$Universit\"at T\"ubingen, 
	 Institut f\"ur Theoretische Physik,\\
         \small \it Auf der Morgenstelle 14,
         D-72076 T\"ubingen, Germany,\\
	 \small \it E-mail: Reinhard.Alkofer@uni-tuebingen.de,
	 chfi@tphys.physik.uni-tuebingen.de\\[3pt]
         \small \it $^{b}$Universit\"at Heidelberg, 
	 Institut f\"ur Theoretische Physik,\\ 
	 \small \it Philosophenweg 16, 
	 D-69120 Heidelberg, Germany}

\maketitle	 


{\small
}

\begin{abstract}
Recent investigations of the Dyson--Schwinger equations and Monte--Carlo lattice
calculations resulted in a coherent description of the fully dressed gluon,
ghost and quark propagators in Landau gauge QCD. In the Dyson--Schwinger
approach the infrared behaviour of these propagators is determined
analytically. For finite spacelike momenta the gluon, ghost and quark
propagators are compared to available corresponding results of lattice
Monte--Carlo calculations. For all three propagators an almost quantitative
agreement is found.
These results for the non-perturbative propagators allow an analytical
verification of the Kugo--Ojima confinement criterion. Our numerical
analysis clearly reveals positivity violation for the gluon propagator 
generated by a cut in the complex momentum plane. 
The non-perturbative strong running coupling resulting from these propagators
possesses an infrared fixed point. The quark propagator obtained from quenched
and unquenched calculations displays dynamical chiral symmetry breaking with
quark masses close to ``phenomenological'' and lattice values. We confirm that 
linear extrapolations of the quark propagator for different bare
masses to the chiral limit are inaccurate.
\end{abstract}

\section{Some Aspects of Confinement}

Data taken at the Jefferson Laboratory over the last few years have uncovered
many unexpected properties of hadrons. It seems that CEBAF operates in a very
interesting energy range, and it is foreseeable that after its upgrade many
more highly interesting results will be obtained.  More or less all of these
investigations aim at an understanding of the structure of hadrons in terms of
the underlying degrees of freedom, the quarks and gluons. To bring our current
theory of strong interactions, Quantum Chromo Dynamics (QCD), into agreement
with observations the hypothesis of confinement is needed. Even thirty years
after the formulation of QCD a detailed understanding of the confinement
mechanism(s) is still lacking.

Experimental data obtained at the Jefferson Laboratory pose a challenge
to theoretical physicists working in this field: In principle the requirement
is to understand hadronic properties at intermediate and large momentum
transfers in terms of QCD degrees of freedom. 
An adequate theoretical approach for such investigations has to
be deeply rooted in Quantum Field Theory to accommodate a description of
confinement, it has to be Poincar{\'e}-covariant to be applicable at the 
employed momentum transfers, and, last but not least, it has to be manageable 
for quite complicated hadronic form factors and reactions.

Non-perturbative calculations of QCD correlation functions do fulfill all 
these requirements. Studies of their equations of motion, the Dyson--Schwinger
equations (DSEs), have the potential to provide a successful phenomenology of hadrons
in terms of quarks and gluons, for recent reviews see {\it e.g.\/} 
\cite{Maris:2003vk,Alkofer:2000wg,Roberts:2000aa}. Within this approach
dynamical breaking of chiral symmetry is obtained from first principles
\cite{Fischer:2003rp}. The description of meson properties and reactions
based on their Bethe--Salpeter amplitudes has been quite successful,
see ref.\ \cite{Maris:2003vk} and references therein, and it seems
viable that covariant Faddeev amplitudes will allow for a computation of
baryonic reactions as well, see {\it e.g.\/} \cite{Ahlig:2000qu}.
On the other hand, the results of ref.\ \cite{Ahlig:2000qu} make clear that
an understanding of confinement and its implications  onto the analytic
structures of QCD Green's functions is a necessary prerequisite for further
progress in this direction.

The phenomenon of confinement  is truely non-perturbative in nature: There is a
physical scale associated with it, and we know from asymptotic freedom of QCD
that such scales are non-analytic in the coupling. Futhermore, as we will see
shortly, QCD Green's functions exhibit infrared singularities related to
confinement. Lattice calculations are very helpful because they provide a
rigorous non-perturbative method, nevertheless, there is definite need for
continuum-based methods.  In the following we will discuss how the infrared
behaviour of QCD propagators can be determined analytically, and what the
corresponding results tell us about confinement in the covariant gauge.

Let us take a step back and discuss Quantum Electro Dynamics (QED) first. In
QED in the covariant gauge the electromagnetic field can be decomposed into
transverse, longitudinal and time-like photons, but the latter two are never
observed. From a mathematical point of view this can be understood from
the representations of the Poincar{\'e} group: Massless
particles have only two possible polarizations. This apparent contradiction is
resolved by the fact that time-like and longitudinal photons cancel exactly in
the $S$-matrix \cite{Bleuler50}. We can interpret this also otherwise:
The time-like photon, being unphysical due to the Minkowski metric from the very
beginning, ``confines'' the longitudinal photon.

In QCD cancelations of unphysical degrees of freedom in the $S$-matrix  also
occur but are more complicated due to the self-interaction of the gluons. One
obtains {\it e.g.\/} amplitudes for the scattering  of two transverse into one
transverse and one longitudinal gluons to order $\alpha_S^2$. A consistent
quantum formulation in a functional integral approach leads to the introduction
of ghost fields \cite{Faddeev67}. To order $\alpha_S^2$ a ghost loop then
cancels all gluon loops which describe scattering of transverse to longitudinal
gluons. The proof of this cancelation to all orders in perturbation theory has
been possible by employing the BRS symmetry of the covariantly gauge
fixed theory~\cite{Becchi:1976nq}. At this point one has achieved a consistent
quantization. If one wants to describe confinement of coloured states 
in a similar efficacious way one has to go only one slight step further:
One has to require that only BRS singlets are allowed as physical states
\cite{Kugo:1979gm,Nakanishi:qm}.

The Kugo--Ojima confinement scenario \cite{Kugo:1979gm,Nakanishi:qm}  describes
a mechanism by which the  physical state space contains only colourless states.
The coloured states are not BRS singlets and therefore do not appear in
$S$-matrix elements: They are confined.  Transverse gluons are BRS-non-singlets
states with gluon-ghost, gluon-antighost and gluon-ghost-antighost states in 
the same multiplet. Gluon
confinement then occurs as destructive interference between these states. In
Landau gauge a sufficient criterion for this type of confinement to occur is
given by the infrared behaviour of the ghost propagator: If it is more singular
than a simple pole the Kugo--Ojima confinement criterion is fulfilled. It is
important to note that the general properties of the ghost DSE
and one additional assumption, namely that QCD Green's functions 
can be expanded in asymptotic series in the infrared, allow to prove this
version of the Kugo--Ojima confinement criterion
\cite{Watson:2001yv,Lerche:2002ep}.

\section{Gluon Propagator and Running Coupling} 

Recently strong arguments have been provided in favor of infrared dominance of
the gauge fixing part of the QCD action in the covariant gauge
\cite{Zwanziger:2003cf}. The related infrared dominance of ghost loops  also
occurs in truncation schemes of DSEs being
self-consistent at the level of two-point functions \cite{vonSmekal:1997is}.
These schemes have been refined and generalized \cite{Fischer:2002hn} and
allowed then to solve the coupled set of DSEs for
the ghost, gluon and quark propagators \cite{Fischer:2003rp}. 

In Landau gauge these  momentum-space propagators
$D_G(p)$, $D_{\mu \nu}(p)$ and $S(p)$ renormalized 
at a scale $\mu$ can be generically written as
\begin{eqnarray}
  D_G(p,\mu^2) &=& - \frac{G(p^2,\mu^2)}{p^2} \,,
  \label{ghost_prop}\\
  D_{\mu \nu}(p,\mu^2) &=& \left(\delta_{\mu \nu} - \frac{p_\mu
      p_\nu}{p^2} \right) \frac{Z(p^2,\mu^2)}{p^2} \, ,
  \label{gluon_prop} \\
  S(p,\mu^2) &=& \frac{1}{-i  p\!\!\!/\, A(p^2,\mu^2) + B(p^2,\mu^2)}
  =  \frac{Z_Q(p^2,\mu^2)}{-ip\hspace{-.5em}/\hspace{.15em}+M(p^2)}
  \, .
  \label{quark_prop}
\end{eqnarray}
Two renormalisation scale independent combinations build
from these functions will be important for the further
discussion: $M(p^2)=B(p^2,\mu^2)/A(p^2,\mu^2)$ denotes the
quark mass function, and  a non-perturbative definition of the
running coupling,
$ \alpha_S(p^2) =
  \alpha_S(\mu^2) \: G^2(p^2,\mu^2) \: Z(p^2,\mu^2)$,
is possible due to the non-renormalisation of the ghost-gluon vertex
in Landau gauge \cite{vonSmekal:1997is}.

Employing asymptotic expansions for the propagators at small momenta
the ghost and gluon equations can be solved analytically.
One finds simple power laws,
\begin{eqnarray}
  Z(p^2,\mu^2) \sim (p^2/\mu^2)^{2\kappa}, \qquad
  G(p^2,\mu^2) \sim (p^2/\mu^2)^{-\kappa},
  \label{g-power}
\end{eqnarray}
for the gluon and ghost dressing function with exponents related to
each other. Hereby $\kappa$ is an irrational number,
$\kappa = (93 - \sqrt{2101})/98 \approx 0.595$
\cite{Lerche:2002ep,Zwanziger:2001kw}.
The product $G^2(p^2,\mu^2)\: Z(p^2,\mu^2)$ goes to a constant in the infrared.
Correspondingly we find an infrared fixed point for the running coupling,
$$
\alpha_S(0)=\frac{4 \pi}{6N_c}
\frac{\Gamma(3-2\kappa)\Gamma(3+\kappa)\Gamma(1+\kappa)}{\Gamma^2(2-\kappa)
\Gamma(2\kappa)} \approx 2.972$$
for the gauge group SU(3).
This result depends slightly on the employed truncation scheme.
Infrared dominance of the gauge fixing part of the QCD action
\cite{Zwanziger:2003cf} implies infrared dominance of ghosts 
which in turn can be used to show \cite{Lerche:2002ep} that $\alpha_S(0)$
depends only weakly on the dressing of the ghost-gluon vertex and not at
all on other vertex functions.

The running coupling as it results from numerical solutions for the gluon,
ghost and quark propagators can be quite accurately fitted by 
the relatively simple function \cite{Fischer:2003rp}
\begin{eqnarray}
\alpha_{\rm fit}(p^2) = \frac{\alpha_S(0)}{1+p^2/\Lambda^2_{\tt QCD}}
+ \frac{4 \pi}{\beta_0} \frac{p^2/\Lambda^2_{\tt QCD}}{1+p^2/\Lambda^2_{\tt QCD}}
\left(\frac{1}{\ln(p^2/\Lambda^2_{\tt QCD})}
- \frac{1}{p^2/\Lambda_{\tt QCD}^2 -1}\right) 
\label{fitB}
\end{eqnarray}
with $\beta_0=(11N_c-2N_f)/3$.
Note that, following ref.\ \cite{Shirkov:1997wi}, 
the Landau pole has been subtracted. 
The scale $\Lambda_{\tt QCD}$ is hereby determined
by fixing the running coupling at a certain scale, {\it e.g.\/}
$\alpha_S (M_Z^2) = 0.118$.

\section{Quark Propagator}

In the quark DSE as well as in the quark loop of the 
gluon equation the  quark-gluon vertex enters. It has proven successful 
\cite{Fischer:2003rp} to assume that the quark-gluon vertex factorizes,
\begin{equation}
\Gamma_\nu(q,k) = V_\nu^{abel}(p,q,k) \, W^{\neg abel}(p,q,k),
\label{vertex-ansatz}
\end{equation}
with $p$ and $q$ denoting the quark momenta and $k$ the gluon momentum. The
non-Abelian factor $W^{\neg abel}$ multiplies an Abelian part  $V_\nu^{abel}$,
which carries the tensor structure of the vertex. For the latter we choose
a construction \cite{Curtis:1990zs} used widely in QED.

The Slavnov--Taylor identity for the quark-gluon vertex implies that the 
non-abelian part
$W^{\neg abel}(p,q,k)$ has to contain factors of the ghost renormalization
function $G(k^2)$. Due to the infrared singularity of the latter the effective
low-energy quark-quark interaction is enhanced as compared to the
interaction generated by the exchange of an infrared suppressed gluon.
Therefore the  effective kernel of the quark DSE contains an integrable
infrared singularity. Further constraints imposed on  $W^{\neg abel}(p,q,k)$ 
are such that the quark mass function is, as required from general
principles, independent of the renormalization point and the one-loop
anomalous dimensions of all propagators are reproduced.

\begin{figure}
\vspace{-0.5cm}
\centerline{
\epsfig{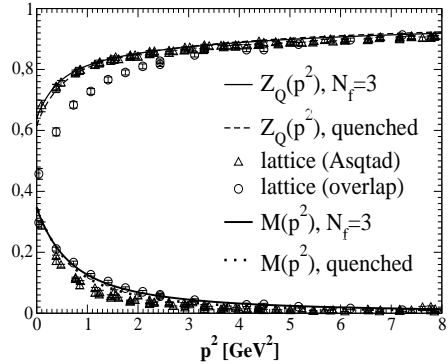}}
\caption{The quark propagator functions in quenched approximation
as well as for three massless flavours 
compared to the lattice data \cite{Bonnet:2002ih}.}
\label{fig:1}
\end{figure}

In Fig.~\ref{fig:1} we compare our results for the  quark propagator
in quenched approximation as well as for three massless flavours with lattice data
\cite{Bonnet:2002ih}. These results nicely agree with the one
from the lattice. Furthermore, for the considered number of flavours the
quenched approximation works well. 

\section{Spectral Properties of the Gluon Propagator}

\begin{figure}
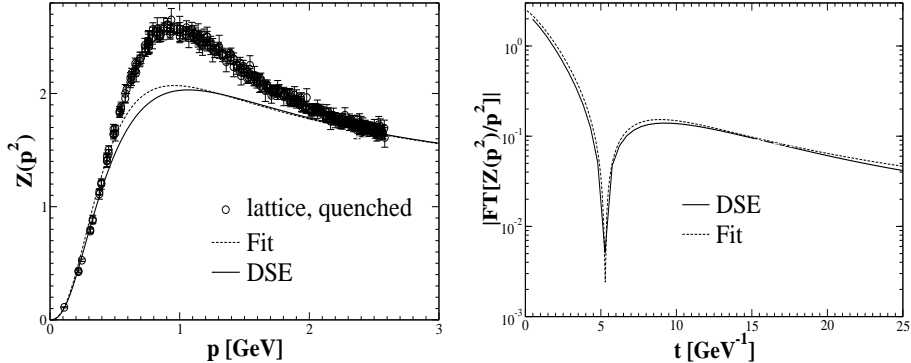

\vspace{-0.5cm}
\centerline{
\epsfig{file=dubro03.glue1.eps,width=5.8cm,height=4.8cm}\hfill
\epsfig{file=dubro03.glue2.eps,width=5.8cm,height=4.8cm}}
\caption{The gluon propagator (left diagram) and its Fourier-transform (right diagram)
compared to the fit, Eq.(\ref{fitII}), and lattice data \cite{Bonnet:2001uh}.}
\label{fig:2}
\end{figure}

The infrared exponent $\kappa$ is an irrational number, and thus  the
gluon propagator possesses a cut on the
negative real $p^2$ axis. It is possible to fit the solution for the gluon
propagator quite accurately without introducing further singularities
in the complex $p^2$ plane.
The fit to the gluon renormalization function \cite{Paper}
\begin{equation}
Z_{\rm fit}(p^2) = w \left(\frac{p^2}{\Lambda^2_{\tt QCD}+p^2}\right)^{2 \kappa}
 \left( \alpha_{\rm fit}(p^2) \right)^{-\gamma}
 \label{fitII}
\end{equation}
is shown in Fig.~\ref{fig:2}. Hereby $w$ is a normalization parameter, and
$\gamma = (-13 N_c + 4 N_f)/(22 N_c - 4 N_f)$
is the one-loop value for
the anomalous dimension of the gluon propagator.  
The corresponding discontinuity along the cut
vanishes for $p^2\to 0^-$, diverges to $+\infty$ at $p^2=-\Lambda_{\tt QCD}^2$  
and goes to zero for $p^2\to \infty$.

The absolute value of the 
Fourier transform of the (transverse) gluon propagator with
respect to Euclidean time is shown in  Fig.~\ref{fig:2}.
First, we clearly observe positivity violations in the gluon propagator.
Second,  the agreement of the numerical Schwinger function with the Fourier
transformed fit is excellent. The
crucial property of the gluon propagator is that it goes to zero for vanishing
momentum.  This can be seen from the relation
$0 = D(p=0) = \int {d^4x} \; D(x)$
(with $D(p) = Z(p^2)/p^2$) which implies that a nontrivial propagator function,
$D(x)$, in coordinate space must contain positive as well as negative
norm contributions.

The function (\ref{fitII}) contains only four parameters:  the overall magnitude
which due to renormalization properties is arbitrary (it
is determined via the choice of the renormalization scale),   the
scale $\Lambda_{\tt QCD}$,   the infrared exponent $\kappa$ and  the
anomalous dimension of the gluon $\gamma$. The latter two are not 
free parameters: $\kappa$ is determined from the infrared properties of the
DSEs and for $\gamma$ its one-loop value is used. Thus we have found a
parameterization of the gluon propagator which has effectively only one
parameter, the scale $\Lambda_{\tt QCD}$. 

\section{Spectral Properties of the Quark Propagator} 

When discussing the results for the quark propagator we stated already
that a dressed quark-gluon vertex is mandatory.   For the present
discussion it is important to note that such or similar solutions of the
Slavnov--Taylor identity for the quark-gluon  vertex will always result
in the appearance of a quark-gluon coupling term $\Delta B$
proportional to the sum of quark momenta,
\begin{eqnarray}
  V_\nu^{abel}(p,q) &:=& \Sigma A_\nu + \Delta B_\nu + ... \nonumber\\
  &=& \frac{A(p^2)+A(q^2)}{2}\gamma_\nu + 
    i\frac{B(p^2,\mu^2)-B(q^2,\mu^2)}{p^2-q^2} (p+q)_\nu + ...\, .
  \label{DeltaB}
\end{eqnarray}
Such a coupling, being effectively scalar, is {\it per se} not invariant
under chiral transformations contrary to the leading term, $\Sigma A$, of the 
quark-gluon vertex. It is important to realize that the term $\Delta B$
appears only in case chiral symmetry is already dynamically
broken. Thus it is consistent with the chiral Ward identities. Its
existence, on the other hand, provides a significant amount of self-consistent
enhancement of dynamical chiral symmetry breaking.
Fairly independently of the form of the gluon propagator the
resulting quark propagator respects positivity if the term $\Delta B$ is
included in the quark-gluon vertex \cite{Paper}.

\begin{figure}
\vspace{-0.5cm}
\centerline{
\epsfig{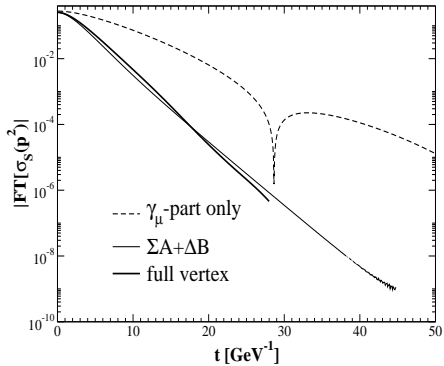}}
\caption{The Fourier transform of the scalar part of the quark propagator
for three different types of vertices. Positivity violations disappear once
the scalar coupling, $\Delta B$, is included in the vertex \cite{Paper}.}
\label{fig:3}
\end{figure}

A peculiar feature of the Fourier transform of the scalar part of the 
quark propagator is the 
curvature appearing for small Euclidean times. There are at least  three
possible sources for this. A leading singularity on the real
momentum axis may be accompanied by additional real singularities at larger
masses or by complex conjugate singularities with a larger real part of the
mass, or it may be the starting point of a branch cut on the negative real
momentum axis \cite{Paper}.

\section{Extrapolation to the chiral limit}

\begin{figure}
\vspace{-0.5cm}
\centerline{
\epsfig{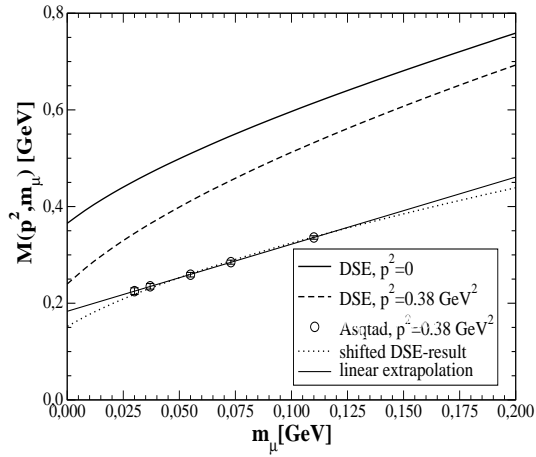}}
\caption{Relation between the dynamically generated quark mass at $p^2=0.38 
\mbox{GeV}^2$ and $p^2=0$ to the renormalized quark mass $m_\mu$
at $\mu=19$ GeV from
the DSEs and from the lattice data of ref.~\cite{Bonnet:2002ih}}
\label{fig:4}
\end{figure}

Lattice simulations of dynamical chiral symmetry breaking are facing problems from 
two sides. Finite volume effects tend to obfuscate results in the very infrared.
Furthermore, although in principle the Ginsparg-Wilson relation enables one to 
implement chiral quarks on a lattice, small quark masses are computationally 
very expensive. Therefore lattice simulations are usually carried out at finite 
bare quark masses and then linearly extrapolated to the chiral limit. The 
DSE-approach, however, suggests that this linear extrapolation is inaccurate 
\cite{Bhagwat:2003vw}. The DSE for the quark propagator also contains the nontrivial 
relation between the dynamically generated quark 
mass at small momenta and the renormalized quark mass at a perturbative 
renormalization point. To illustrate this point we display the dynamical mass 
$M(p^2,m_\mu)$ from the DSEs for two different momenta, $p^2=0$ and 
$p^2=0.38 \,\mbox{GeV}^2$ in Fig.~\ref{fig:4}. One clearly observes 
the curvature in the DSE-results for both momenta, reflecting the nonlinear behaviour 
of the underlying equations. We compare these results with lattice data taken from 
ref.~\cite{Bonnet:2002ih} and $m_\mu$ from 
ref.~\cite{Bhagwat:2003vw}. The lattice data are consistent with a 
linear fit. However, by multiplying our DSE-results with an overall factor we 
obtain a curved mass function that mimics a DSE-model-fit to the lattice data.
The extrapolated chiral dynamical quark mass at $p^2=0.38 \, \mbox{GeV}^2$ is roughly
20\% below the one obtained from the linear fit. Employing only the leading
$\gamma_{\mu}$-structure of the quark-gluon vertex, as done in ref.~\cite{Bhagwat:2003vw},
leads to even more drastic deviations. Thus we confirmed that the linear extrapolation
of lattice data to the chiral limit cannot be trusted well.

\section{Epilogue}

Recent years have seen a lot of progress in Strong QCD. In Landau gauge
we gained an understanding of the infrared behaviour of gluon and 
quark propagators. We have verified the Kugo--Ojima scenario for gluon
confinement and found dynamical chiral symmetry breaking in an {\it ab initio}
calculation.

We have proposed relatively simple functions to describe the running coupling,
the gluon and the quark propagators (see ref.~\cite{Paper} for details) for 
all possible values of momenta. These have the potential to provide a basis 
for a hadron phenomenolgy based on quarks and gluons, even and especially 
in the non-perturbative regime.

\section*{Acknowledgments}

R.\ A.\ thanks the organizers for their support. 
He is especially thankful to Dubravko Klabucar for his kindness and hospitality.  

\noindent
We are grateful to P.~van Baal, P.~Bowman, W.~Detmold, H.~Gies,
P.~Maris, C. Roberts,
I.~Solovtsov, O.~Solovtsova,
P.~Tandy, P.~Watson, A.~Williams and D. Zwanziger
for many helpful hints and enlightening discussions.

\noindent
This work was supported by DFG under contracts Al 279/3-4, Gi 328/1-2 
and GRK683 (European graduate school Basel--T\"ubingen).

\end{document}